\documentclass[10pt,dvips]{article}
\usepackage{amssymb,amsfonts,amsmath,latexsym}
\usepackage{graphics,graphicx,epsf}
\newcommand{\be}{\begin{equation}}
\newcommand{\ee}{\end{equation}}
\newcommand{\bea}{\begin{eqnarray}}
\newcommand{\eea}{\end{eqnarray}}
\newcommand{\ba}{\begin{array}}
\newcommand{\ea}{\end{array}}
\newcommand{\bt}{\begin{tabular}}
\newcommand{\et}{\end{tabular}}

\newcommand{\fr}{\frac}
\newcommand{\ci}{\cite}
\newcommand{\cl}{\centerline}
\newcommand{\bs}{\bigskip}

\newcommand{\vs}{\vspace}

\newcommand{\en}{\eqno}

\newcommand{\fns}{\footnotesize}

\newcommand{\bbib}{\begin{thebibliography}}
\newcommand{\ebib}{\end{thebibliography}}

\newcommand{\und}{\underline}

\begin{document}
\bs
\cl{\bf DUALITY AND EXACT RESULTS FOR CONDUCTIVITY} \cl{\bf OF 2D
ISOTROPIC HETEROPHASE SYSTEMS}
\cl{\bf IN MAGNETIC FIELD}

\bs

\cl{{\bf S.A.Bulgadaev} \footnote{e-mail: bulgad@itp.ac.ru}, {\bf
F.V.Kusmartsev} \footnote{e-mail: F.Kusmartsev@lboro.ac.uk}}

\bs
\cl{\fns Landau Institute for Theoretical Physics,
Chernogolovka, Russia, 142432}
\cl{\fns Department of Physics,
Loughborough University,  Loughborough, LE11 3TU, UK}

\bs

\begin{quote}
\footnotesize{ Using a fact that  the effective conductivity $\hat
\sigma_{e}$ of 2D random heterophase systems in the orthogonal
magnetic field is transformed under some subgroup of the linear
fractional group, connected with a group of linear transformations
of two conserved currents, the exact values for $\hat \sigma_{e}$
of isotropic heterophase systems are found. As known, for binary
(N=2) systems  a determination of exact values of both
conductivities (diagonal $\sigma_{ed}$ and transverse Hall
$\sigma_{et}$) is possible only at equal phase concentrations and
arbitrary values of partial conductivities. For heterophase ($N
\ge 3$) systems this method gives exact values of effective
conductivities, when their partial conductivities belong to some
hypersurfaces  in the space of these partial conductivities and
the phase concentrations are pairwise equal. In all these cases
$\sigma_e$ does not depend on phase concentrations. The complete,
3-parametric, explicit transformation, connecting $\sigma_e$ in
binary systems with a magnetic field and without it, is
constructed.}
\end{quote}

\bs
\cl{PACS: 75.70.Ak, 72.80.Ng, 72.80.Tm, 73.61.-r}
\bs

\underline{1. Introduction}

\bs
The properties of the electrical transport of
classical macroscopically inhomogeneous (random or regular) heterophase systems,
consisting of $N \;(N \ge 2)$  phases with different
conductivities  $\sigma_i \; (i =1,2,...,N),$ have had always an importance for practice \ci{1}.
Last time new materials appeared which are very perspective from the high technology point of view.
They often have the heterophase inhomogeneous structure on smaller scales
(from microscopic till mesoscopic) and some unusual transport properties. For example,
the magnetoresistance of oxide materials with a perovskite type  structure
becomes very large (the so called colossal magnetoresistance in such materials as manganites) \ci{2}
or grows approximately linearly with magnetic field up to very high fields (in silver chalcogenides) \ci{3}.
There is an opinion that these properties take place due to inhomogeneities  of these materials \ci{3}.
For this reason a calculation of the effective conductivity $\sigma_{e}$ of inhomogeneous
heterophase systems without and with magnetic field at arbitrary partial conductivities and
phase concentrations is very important problem. Unfortunately, the existing effective medium
approximations (EMA) describes well the effective conductivity of random heterophase systems
without magnetic field only for binary ($N=2$) systems and at not too large inhomogeneities \ci{4}.
For systems with $N>2$ phases without magnetic field \ci{4,5} and for systems in magnetic field \ci{6}
the EMA cannot give an explicit simple formulas convenient for description of the experimental results
in a wide range of partial parameters. Such formulas can be obtained only in high magnetic field \ci{6,7}

A situation is better in 2D systems, where a few exact results
\ci{8,9,10} and approximate explicit expressions  based on exact
duality relations (DR)  have been obtained for isotropic self-dual
heterophase systems without magnetic field \ci{5,11}. But, the
exact results for inhomogeneous systems in magnetic field have
been obtained only for binary systems at equal phase
concentrations $x_1 = x_2 =1/2$ \ci{12,13} (see also \ci{14} about systems with
uniform $\sigma_t$). In this letter we will
show that the exact results for both diagonal ($\sigma_{ed}$) and
transverse ($\sigma_{et}$) effective conductivities can be
obtained for isotropic self-dual heterophase systems (random as
well as regular) in  magnetic field $H$ and at arbitrary number of
phases $N$. They generalize the known results for binary systems
with magnetic field \ci{12,13} and for arbitrary $N$-phase systems
without magnetic field \ci{10}. In particular, it will be shown
that for $N>2,$ contrary to the $N=2$ case, there exist the whole
hyperplanes in the space of concentrations and hypersurfaces in
the space of partial conductivities, on which $\sigma_e$ does not
depend on phase concentrations. The full explicit transformation,
connecting $\sigma_e$ in systems with a magnetic field and without
it, is constructed.

\bs
\underline{2. Conserved currents, symmetries and duality transformations}

\und{in magnetic field}

\bs

We start with a brief description of some general properties of
classical isotropic stationary conducting systems in the external
perpendicular magnetic field $H$. Such systems are described by
the following differential equations for a current ${\bf j}({\bf
r})$ and an electric field ${\bf e}({\bf r})$ \ci{1}
$$
{\bf div} \cdot {\bf j}({\bf r}) = \partial_i j_i({\bf r})=0,
\quad \nabla \times {\bf e}({\bf r}) = \epsilon_{ik}
\partial_i e_k =0, \quad i,k =1,2 \en(1)
$$
and the Ohm law
$$
{\bf j}({\bf r}) = \hat \sigma ({\bf r}) {\bf e}({\bf r}),
\en(2)
$$
where
$$
\hat \sigma = \sigma_{ik} = \sigma_d \delta_{ik} + \sigma_t \epsilon_{ik}, \quad
\sigma_d ({\bf H}) = \sigma_d (-{\bf H}), \quad
\sigma_t ({\bf H}) = -\sigma_t (-{\bf H}),
\en(3)
$$
All conductivities $\hat \sigma$ of the form (3) commute between themselves
and can be treated as ordinary functions.
The effective conductivity $\hat \sigma_{e}$ of $N$-phase random
or regular symmetric systems with the partial conductivities
$\sigma_{id}, \; \sigma_{it} \;(i = 1,2,...,N)$ (we assume that
$\sigma_{id} \ge 0$) and concentrations $x_i$ (satisfying the
normalization condition $\sum_{i=1}^N x_i =1$) must be a symmetric
function of pairs of arguments ($\hat \sigma_i, x_i$) and a
homogeneous (a degree 1) function of $\sigma_{di,ti}.$ For this
reason it is invariant under all permutations
$$
\hat \sigma_{e}(\hat \sigma_1, x_1|\hat \sigma_2, x_2|...|\hat
\sigma_N, x_N) = \hat \sigma_{e}(P_{ij}(\hat \sigma_1, x_1|\hat
\sigma_2, x_2|...|\hat \sigma_N, x_N)),
\en(4)
$$
where $P_{ij}$ is the permutation of the i-th and j-th pairs of arguments,
and can be represented in the next form
$$
\hat \sigma_{e}(\hat \sigma_1, x_1|\hat \sigma_2, x_2|...|\hat
\sigma_N, x_N) = \sigma_s \hat
f(z_{1d,1t},x_1|z_{2d,2t},x_2|...|z_{Nd,Nt},x_N). 
$$
Here $z_{id,it} = \sigma_{id,it}/\sigma_s,$ where $\sigma_s$ is some normalizing
conductivity. It is convenient to choose $\sigma_s$
in the  form symmetrical relative $\sigma_{id,it},$ what conserves a symmetry property of $f.$

The effective conductivity of $N$-phase systems must satisfy, together with the usual boundary
conditions on boundaries between i-th and k-th phases
$$
j_{in} = j_{kn}, \quad e_{i||} =e_{k||}, \quad i,k =1,2,...,N,
\en(5)
$$
the following natural reduction and limiting requirements:

1) $\hat \sigma_e$ of $N$-phase system with
$n \; (2 \le n \le N)$ equal partial conductivities must reduce to
$\hat \sigma_e$ of system with $N-n+1$ phases and
the concentrations of the phases with equal conductivities must add;

2) it has not depend on partial $\sigma_{di,ti}$ and must reduce to the effective
conductivity of $(N-1)$-phase system, if the concentration of this phase
$x_i = 0$;

3) it must reduce to some partial $\hat \sigma_i$, if $x_i = 1.$

The duality relations for $\hat \sigma_{e}$ in magnetic field have
in general a more richer and complicated structure, which
generalizes the inversion structure of DR of systems without
magnetic field.  It was established firstly by Dykhne in \ci{12},
where an idea and a method to solve this problem have been
proposed. We use these idea and method, refined later (see, for example, \ci{13,15,16}), in a modern form. The main observation is that in the 2D conducting
system, described by equations (1,2), really exist 2 conserved
currents and two curlfree vector fields. The second partners for
${\bf j}$ and ${\bf e}$ are their conjugated fields, which satisfy
the corresponding equations
$$
\tilde j_i = \epsilon_{ik} e_k, \quad \tilde e_i = \epsilon_{ik} j_k, \quad
\partial_i \tilde j_i =0, \quad \epsilon_{ik} \partial_i \tilde e_k =0.
\en(6)
$$
As a result one can consider  linear combinations of conserved
currents and curl-free electric fields
$$
{\bf j} = a {\bf j}' + b \tilde {\bf j}', \quad
{\bf e} = c {\bf e}' + d \tilde {\bf e}',
\en(7)
$$
where the coefficients $a,b,c,d \in R$ are arbitrary real. Futher it will be more convenient to use
the complex representation for coordinates and vector fields \ci{1}
$$
z=x+iy, \quad j= j_x+ij_y, \quad e=e_x+ie_y,  \quad \sigma = \sigma_d + i\sigma_t.
$$
The primed currents and electric field
satisfy again the Ohm law
$$
 j' = \sigma'  e', \quad  \sigma' = T(\sigma) = \fr{c \sigma -ib}{-id \sigma + a}.
\en(8)
$$
It follows from (7),(8) that under a linear transformation (7) a conductivity transforms under
the corresponding linear fractional (LF) transformation T from (8). In the absence of magnetic field $\sigma$ is real
(or, in a matrix representation, diagonal) and a transformation (8) reduces to the  inversion  \ci{8,9}
$$
\sigma' = \fr{\sigma_0^2}{\sigma}, \quad \sigma_0^2 = b/c, \quad a=d =0
\en(9)
$$
Thus, instead of one-parametric inversion transformations in
systems without magnetic field, $\sigma$ in magnetic field
transforms under linear fractional transformation with 3 real
parameters (since one of 4 parameters can be factored due to the
fractional structure of T). There are various ways to choose 3
parameters. In our treatment
it will be convenient to factor $d.$ This gives 3 parameters
$\bar a = a/d, \bar b= b/d,\bar c= c/d,$ determining a
transformation $T.$ The transformations $T$ from (8) form a subgroup of
a group of all linear fractional transformations with arbitrary
complex coefficients $a,b,c,d,$ conserving the imaginary axis.
This subgroup contains also the shift transformations for an
imaginary part of $\sigma$, which correspond to $d=0, \; a=c.$

The transformation (8) has the following form in terms of conductivity components $\sigma_d$ and $\sigma_t$
$$
\sigma_d' = \sigma_d \fr{ac + bd}{(d \sigma_d)^2 + (a+
d \sigma_t)^2} = \bar c \sigma_d \fr{\bar a  + {\bar b}/{\bar c}}{( \sigma_d)^2
+ (\bar a + \sigma_t)^2},
$$
$$ \sigma_t' = \fr{cd \sigma_d^2 + (a+d
\sigma_t)(c\sigma_t -b)}{(d \sigma_d)^2 + (a+ d \sigma_t)^2} = {\bar
c} \fr{ \sigma_d^2 + ({\bar a} + \sigma_t)(\sigma_t -b/c)}{(
\sigma_d)^2 + ({\bar a}+ \sigma_t)^2}. \en(10)
$$
They have rather interesting structure. Any
general transformation gives nontrivial transverse part
$\sigma_t'\ne 0$, even if it was absent ($\sigma_t = 0$), while
$\sigma_d'$ remains always equal to $0,$ if $\sigma_d=0$ (or
$\sigma_d'=0,$ if det $T =0$).

This general consideration can be applied also to the effective conductivity of inhomogeneous systems.
Then one obtains the following transformation property for $\sigma_e$
$$
\sigma_e(\{\sigma_i\},\{x_i\}) = T (\sigma_e( T (\{\sigma_i\}),\{x_i\})),
$$
$$
T^{-1} (\sigma_e(\{\sigma_i\},\{x_i\})) = \sigma_e( T (\{\sigma_i\}),\{x_i\})).
\en(11)
$$
The relation (11) means that $\sigma_e$  transforms under
conjugated representation ($\sim T^{-1}$) of linear fractional
transformations and its change as a function of its arguments can
be completely compensated by the same transformation of it itself
(this form takes place for the dual transformations connected with symmetry 
transformations satisfying the condition $T \sim T^{-1}$ and for them 
is equivalent to the usual case from \ci{12}).
Note also that the transformations (11) does not depend and does
not act on concentrations. Following a tradition, we will call
such transformations the duality transformations (DT) and such
relations the duality relations (DR). An existence of the such
exact duality relations is very important property of
two-dimensional inhomogeneous systems. For example, they can be
used for the obtaining the exact values of effective conductivity.
For this one needs the transformations, which action on the set of
partial conductivities, can be compensated by symmetries of the
effective conductivity as a function of its partial conductivities
and phase concentrations. It means that the set of all partial
parameters and concentrations must be invariant under some joint
action of the DT and symmetry operations, i.e. it must be their
fixed point (FP).   At this set of partial parameters $(\{\sigma*_i\}, \{x*_i\})$ one obtains for $\sigma_e$ the exact equation
$$
\sigma_e(\{\sigma*_i\},\{x*_i\}) = T
(\sigma_e(\{\sigma*_i\},\{x*_i\})), \en(12)
$$
where $T$ is the corresponding DT. This equation means also that
$\sigma_e$ at the FP is a fixed point (or an eigenvector with the eigenvalue equal to 1) of $T.$ In order to exclude a misunderstanding of different fixed points, we will call the fixed points of $T$ as its eigenvectors.
As is known, the LF transformations  $T$  can have only  one or
two eigenvectors in dependence on value of its discriminant. For
$T$ from (8) one has for such eigenvectors $z_{1,2}$
$$
T(z_{1,2}) = z_{1,2}, \quad z_{1,2} = -i\fr{a-c}{d} \pm \sqrt{\fr{-(a-c)^2+ 4bd}{4d^2}}
\en(13)
$$
It follows from (13), that for $(a-c)^2< 4bd$ the corresponding $z_{1}$ will have a positive real part, what gives a physical conductivity. For $(a-c)^2> 4bd$ both eigenvectors have only imaginary part, i.e.
they can correspond only to $\sigma$ with $\sigma_d =0,\; \sigma_t \ne 0.$ In classical and usual systems this case is unphysical and cannot be realized (though, it can serve for a phenomenological approximate description of
the low-temperature systems with quantum Hall effect \ci{15}). It follows from (13) that one can always construct for a given physical $\sigma$ the transformation $T_{\sigma},$ which has $\sigma$ as its eigenvector,  
$$
\sigma = \sigma_d + i \sigma_t, \quad
2\sigma_t = - (\fr{a-c}{d}),  \quad b/d = \sigma_d^2 + \sigma_t^2,
\en(14)
$$
As one can see from (14), this $\sigma$ does not define the corresponding $T_{\sigma}$ unambiguously, since there are 3 real parameters in $T.$
The third parameter can be determined if $T$ has some additional constraint. There is a very important case of
transformations $T,$ satisfying the condition $T^2(z) = z,$ (i.e. $T^2 \sim I$ or
$T^{-1} \sim T,$ it includes all $T$ acting as permutations). For these transformations $a=-c$ and, consequently, one has
$$
\sigma_t = - a/d,  \quad b/d = \sigma_d^2 + \sigma_t^2.
\en(15)
$$
For transformations with $c=a$ and $b/d \ge 0$ the eigenvectors can be only real $z_{1,2} = \pm \sqrt{b/d}.$

Thus, we see that though a number of parameters of "dual" transformations $T$ becames larger (3 instead of 1), but, since a number of partial conductivities doubled, they are still not enough to make the DR essentially richer, than in a case with $H=0$ (netherless, some new possibilities appear,
see below section 4). Fortunately, in any case one can define the transformation $T_{12},$ interchanging arbitrary $\sigma_1$ and $\sigma_2,$
$$
T_{12} \sigma_1 = \sigma_2, \quad T_{12} \sigma_2 = \sigma_1
\en(16)
$$
In components (16) has the form
$$
\sigma_{2d} =  \fr{\bar c \sigma_{1d}(\bar a + \bar b/\bar c)}{(\sigma_{1d})^2 + (\bar a +
\sigma_{1t})^2}, \quad \sigma_{2t} = \bar c \fr{\sigma_{1d}^2 + (\bar a +
\sigma_{1t})(\sigma_{1t} -b/c)}{\sigma_{1d}^2 + (\bar a +
\sigma_{1t})^2}
$$
$$
\sigma_{1d} = \fr{c \sigma_{2d}(\bar a + b/c)}{(\sigma_{2d})^2 + (\bar a +
\sigma_{2t})^2}, \quad \sigma_{1t} = \bar c \fr{\sigma_{2d}^2 + (\bar a +
\sigma_{2t})(\sigma_{2t} -b/c)}{(\sigma_{2d})^2 + (\bar a +
\sigma_{2t})^2}. \en(17)
$$
Though, formally, this transformation relates 4 independent parameters and satisfies the condition $T^2 \sim I,$ it has one-parametric general solution of the form
$$
a= -d \fr{\sigma_{1d} \sigma_{2t} + \sigma_{2d} \sigma_{1t}}{\sigma_{1d} + \sigma_{2d}},
\quad c =-a,
$$
$$
b = d \fr{\sigma_{1d} \sigma_{2t}^2 + \sigma_{2d} \sigma_{1t}^2 + \sigma_{1d} \sigma_{2d}(\sigma_{1d} + \sigma_{2d})}{\sigma_{1d} + \sigma_{2d}},
\en(18)
$$
where $d$ is included as an arbitrary common factor. This fact reflects an existence of some hidden symmetry, wich effectively reduces a number of independent equations (see also discussions in the Sections 3 and 4).

\bs

\und{3. Fixed points and exact values of the effective conductivity.}
\bs

Having all mathematical machinery ready, let us consider the
nontrivial fixed points of the DT. We begin with the known results
for binary case $N=2.$

(1). A binary case $N=2.$

In this case the effective conductivity $\sigma_e$ depends on two sets of partial parameters
$\sigma_{id}, \sigma_{it}, x_i, i=1,2.$ As it was shown in the previous section, one can always find
the DT $T_{12}$, interchanging  partial conductivities of two phases. Then, using (11) and a symmetry of
$\sigma_e$ one obtains the following DR
$$
\sigma_e(\sigma_1,x_1|\sigma_2,x_2) = T_{12} (\sigma_e(\sigma_1,x_2|\sigma_2,x_1)),
\en(19)
$$
At the point of equal phase concentrations one obtains the exact equation for $\sigma_{e(12)}$
$$
\sigma_e(\sigma_1,1/2|\sigma_2,1/2) = T_{12} (\sigma_e(\sigma_1,1/2|\sigma_2,1/2)),
\en(20)
$$
which means that this $\sigma_e$ is a fixed point (or an
eigenvector) of $T_{12}$ transformation. From (15) and (18) one
gets the exact value for $\sigma_e$
$$
\sigma_{ed}(\sigma_{1d},1/2|\sigma_{2d},1/2) = \sqrt{-\bar a^2 +\bar b} =
\sqrt{\sigma_{1d} \sigma_{2d}}\left(1+ \left(\fr{\sigma_{1t} - \sigma_{2t}}{\sigma_{1d} + \sigma_{2d}}\right)^2\right)^{1/2},
$$
$$
\sigma_{et}(\sigma_1,1/2|\sigma_2,1/2) = -\bar a =
\fr{\sigma_{1d} \sigma_{2t} + \sigma_{2d} \sigma_{1t}}{\sigma_{1d} + \sigma_{2d}},
\en(21)
$$
which firstly has been obtained in this form in \ci{13}. It is
easy to see that the change ${\bf H} \to - {\bf H}$ changes a sign
of $\sigma_{et}$ only, in accordance with the Onsager symmetry
(3). The exact formulas (21) are very interesting result. It shows
that the change of the exact result for $\sigma_{ed}$ at $H=0$ is
completely determined by a difference of the partial transverse
conductivities $\sigma_{1t} - \sigma_{2t}$. When they are equal,
the diagonal part of the effective conductivity coincides (in its
form) with that without magnetic field and the transverse part of $\sigma_e$
remains unchanged. Later it was shown that this property takes place for all concentrations,
what allows to describe the systems with uniform transverse conductivity
in terms of systems without a magnetic field \ci{14}.
At the same time, when $\sigma_{1d} =
\sigma_{2d},$ the effective transverse conductivity is simply
equal to the average of partial transverse conductivities
$\sigma_{et} =(\sigma_{1t} + \sigma_{2t})/2$.

(2). 3 phase self-dual systems.

For 3-phase systems without magnetic field the FPs $(\{\sigma*\}, \{x*\})$ of the DT exist, when
two any phases have equal concentrations, for example, $x_1=x_2=x,$(but otherwise they are arbitrary, except the normalization condition $2x+x_3 =1$) and their partial conductivities satisfy the condition
\ci{9,10}
$$
x_1=x_2, \quad 2x+x_3 =1; \quad \sigma_3 = \sqrt{\sigma_1
\sigma_2}. \en(22)
$$
One can try to find the FP of the same form for 3-phase systems
with magnetic field. Then, the interchanging transformation
$T_{12}$ can again be compensated by the permutation symmetry due
to the equality $x_1=x_2$, and one obtains that the FP of this
type can exist if
$$
T_{12} \sigma_1 = \sigma_2, \quad T_{12} \sigma_2 = \sigma_1, \quad
T_{12} \sigma_3 = \sigma_3.
\en(23)
$$
The last condition means that $\sigma_3$ must be also an eigenvector of $T_{12}$ transformation. It is possible only if $\sigma_3$ coincides with the effective conductivity  for two phase systems at equal
phase concentrations $\sigma_e^{(N=2)}$ from (21) (since this $\sigma_e^{(N=2)}$ is also the eigenvector of $T_{12}$)
$$
\sigma_{3d} = \sqrt{\sigma_{1d} \sigma_{2d}}\left(1+
\left(\fr{\sigma_{1t} - \sigma_{2t}}{\sigma_{1d} +
\sigma_{2d}}\right)^2\right)^{1/2}, \quad
\sigma_{3t} =
\fr{\sigma_{1d} \sigma_{2t} + \sigma_{2d} \sigma_{1t}}{\sigma_{1d}
+ \sigma_{2d}}. \en(24)
$$
The equalities (24) generalize the corresponding equalities (22) for systems with $H=0$ and reduce to them when $\sigma_{it} = 0, \; i=1,2,3.$
Thus, we obtain that the effective conductivity of 3-phase self-dual systems for the set of partial parameters $(\{\sigma*\}, \{x*\})$, satisfying (24) and $x_1 = x_2,$, is also the eigenvector of $T_{12}$ and, consequently, coincides with two phase effective conductivity from (21)
$$
\sigma_e^{(N=3)}(\{\sigma*\}, \{x*\})  = \sigma_3 = \sigma_{e(12)}^{(N=2)}.
\en(25)
$$
The set of partial parameters, satisfying the FP conditions, has the same structure as  the corresponding
sets for 3-phase systems without  magnetic field and generalizes it. In this case $\sigma_e$ does not
depend on common concentration of two first phases 
and system behaves as a homogeneous one.

The similar FPs exist, when phase concentrations
of other pairs of phases are equal:
$$
1)\quad  x_1 = x_3, \quad \sigma_2 = \sigma_{e(13)}^{(2)}, \quad 2)
\quad x_2 = x_3, \quad \sigma_1 = \sigma_{e(23)}^{(2)}, \en(26)
$$
No any other nontrivial FPs appear. For example, a new FP can appear, when
all concentrations are equal $x_1=x_2=x_3 =1/3.$ In this case, in principle, it is possible that
the phases are permutated along the shemes (nontrivial cycles)
$"1" \to "2" \to "3" \to "1"$ (or $"1" \to "3" \to "2" \to "1").$
It means that a transformation $T$ must exist  such that
$$
T (\sigma_1) = \sigma_2, \quad T (\sigma_2) = \sigma_3, \quad  T (\sigma_3) = \sigma_1, \quad T^3 \sim I.
\en(27)
$$
Let us consider the corresponding transformations of the diagonal components
$$
\sigma_{2d} =  \fr{\bar c \sigma_{1d}(\bar a + b/c)}{(\sigma_{1d})^2 + (\bar a+
\sigma_{1t})^2}, \quad \sigma_{3d} =  \fr{\bar c \sigma_{2d}(\bar a + b/c)}{
\sigma_{2d}^2 + (\bar a+ \sigma_{2t})^2},
$$
$$
\sigma_{1d} =  \fr{\bar c \sigma_{3d}(\bar a + b/c)}{(\sigma_{3d})^2 + (\bar a+
\sigma_{3t})^2}.
\en(28)
$$
It appears that the condition $T^3 \sim I$ determines $T$ up to one parameter, say $a$
$$
a=c, \quad b/d = 3a^2/d^2.
\en(29)
$$
Then, one obtains from (28) for $a$ 3 relations, which contain 6
parameters (partial conductivities). Another 3 independent
relations can be obtained analogously from expressions for
transverse components (they are more complicated, for this reason
we do not write them here). As a results one has 6 various
expressions for $a.$ They give 15 nonlinear relations for 6
parameters. In principle, this overdetermined system of relations
can have nontrivial solutions, even parametric, if there is some
hidden symmetry, which reduces number of independent relations.
Such situation usually takes  place for exactly solvable models
\ci{17}. If any solutions of this system of relations  exist, they
will give only diagonal eigenvectors, since, due
to $\bar a = \bar c,$ the corresponding eigenvectors of this T  are
$z_{1,2} =\pm \sqrt{3}|\bar a|.$ This means that at this FP
$\sigma_{et} = 0.$ Such unusual result, though it seems an impossible one,
needs a separate consideration.

\bs

(3). General $N$-phase case
\bs

We show now that for heterophase systems with arbitrary number $N$ of phases in magnetic field
the duality relation (4) admits also the FPs, which are a generalization of the FP of the $N$-phase systems
with $H=0$.

Of course, for systems with $N > 3,$ there are the  fixed points with a structure similar to the $N=3$ case,
when for some pair of phases $(i,j)$ their concentrations are equal $x_i = x_j$ and all other
$\sigma_l \; (l\ne i,j)$ are equal between
themselves and are equal to $\sigma_e$ of these two phases. But this case is trivial, since it reduces
to the 3-phase case.
Another possibilities appear for $N>3.$
It will be convenient to consider the cases with even and odd $N$ separately.
Let us consider firstly a case when $N$ is even $N=2M.$
Then a new fixed point is possible, which
corresponds to $M$ pairs of phases with equal concentrations
and to the same transformation $T_{12}$ from (16). But now, $T_{12}$ must act as the interchanging operator on all $M$ pairs.
For example, the fixed point with the phase concentrations
$$
x_{2i-1}=x_{2i}=y_i, \quad \sum y_i =1/2,
 \quad (i=1,...,M)
\en(30)
$$
is possible, if
$$
T_{12}\sigma_{2i-1} = \sigma_{2i}, \quad T_{12}\sigma_{2i} = \sigma_{2i-1},
\quad (i=1,...,M)
\en(31)
$$
For this one needs that all elements $a,b,c,d$ of $T_{12}$ must have the same dependence on the
partial parameters of all pairs of phases
$$
a= -d \fr{\sigma_{(2i-1)d} \sigma_{2it} + \sigma_{2id} \sigma_{(2i-1)t}}{\sigma_{(2i-1)d} + \sigma_{2id}},
\quad c =-a,
$$
$$
b = d \fr{\sigma_{(2i-1)d} \sigma_{2it}^2 + \sigma_{2id} \sigma_{(2i-1)t}^2 + \sigma_{(2i-1)d} \sigma_{2id} (\sigma_{(2i-1)d}+ \sigma_{2id})}{\sigma_{(2i-1)d} + \sigma_{2id}},
\quad i= 1,2...M.
\en(32)
$$
Here  partial conductivities of all phase pairs must be different. Then the equalities (32) define
surfaces in the 4-dimensional spaces of partial parameters of the phase pairs.
The effective conductivity $\sigma_e$ of $N$-phase system has in this case the exact value
$$
\sigma_{e} (\{\sigma\}, \{x\}) = \sigma_e^{(2)} = \sigma_{ei}^{(2)}, \quad
(i=1,...,M),
\en(33)
$$
where $\sigma_{ei}^{(2)}$ is the effective conductivity of the i-th pair, which all are equal.
Note that again the equal concentrations of the phase pairs $y_a \; (a=1,...,M)$
can be arbitrary, except the normalizing condition $\sum_{a=1}^M 2y_a =1.$
The exact value (31) does not depend on $y_a$ and ensures again the correct
values for $\sigma_e$ in the limits $y_a \to 0, \; 1 \; (a=1,...,M).$
Of course, the similar fixed points exist for other ways of the partition
of $N$ phases into the pairs  with equal concentrations.
A number of such points is equal $\#_{2M} = (2M-1)!!.$
The exact values of $\sigma_e$ in these points have always the same general
form
$$
\sigma_{e}^{(N)}(\{\sigma*\}, \{x*\}) = \sigma_{e(ij)}^{(2)} \quad
(i<j, \; i,j = 1,...,N),
$$
where possible pairs $(ij)$ correspond to the phases with equal
concentrations in the corresponding fixed points.

Now we consider a case when $N$ is odd $N=2M+1.$ In this case a new fixed
point is possible if the above conditions of the even case (29) are
supplemented by a condition on $\sigma_{2M+1}$ analogous to the
condition for $N=3$ case
$$
T_{12} \sigma_{2M+1} =   \sigma_{2M+1},
\en(34)
$$
which ensures again the equality of $\sigma_{2M+1}$ with $\sigma_e$ of the whole system. Thus, $\sigma_e$ is again equal to the exact value (21),
coinciding with the exact value of system with $N=2M$.
Of course, the other fixed points related with various ways of choice of
the phase pairs with equal concentrations are possible.
Their number is equal to
$\#_{2M+1} = (2M+1)!!.$
A general form of the exact values of $\sigma_e$ remains the same as in the
even case with an additional equality
$$
\sigma_e^{(N)} (\{\sigma*\}, \{x*\}) = \sigma_{e(ij)}^{(2)} = \sigma_{k}, \quad
i\ne j \ne k, \;\; i <j \quad (i,j,k = 1,...,N).
\en(35)
$$
where the pairs $ij$ correspond to the phases with equal
concentrations and $k$ corresponds to the unpaired phases. Note
that the structure of FPs is in an accordance with the symmetry.

 It follows also from analysis of the case $N=3$ that the exact values
of $\sigma_e$ at the point, where all phase concentrations are
equal, are not universal, when partial conductivities have general
values. An existence of new FP for $N>3$ connected with higher
nontrivial  cycles is also improbable, since all equations of
the form $T^n \sim I, \; (n>3)$ require $a=c$  and the corresponding FPs can give only diagonal $\sigma_e$ (but they also need an additional consideration).

\bs
\und{4. Transformation between systems with ${\bf H}\ne 0$ and ${\bf H}= 0$}
\bs

A more richer structure of the DT in systems with magnetic field
allows also to construct various transformations, connecting
effective conductivities of the initial system and other systems,
related with the initial one. For example, the transformations,
connecting effective conductivities of the systems in opposite
magnetic fields, were constructed \ci{12,15}. Moreover, the
transformation, connecting effective conductivities of two-phase
systems with magnetic field and without it was constructed
\ci{16}. Strictly speaking, it connects the conductivities of
systems with nonzero transverse component and systems with a pure
diagonal structure. Such transformation allows to find the
effective conductivity of inhomogeneous systems in a magnetic
field, when it is known for these systems without a magnetic
field. But, a derivation of this transformation in \ci{16} was
incomplete (in particular, it was not checked for known exact values (21)). 
Under its derivation two simplifying conjectures have been used: (1) the
transformation, symmetrizing partial conductivities $\sigma_1$
and $\sigma_2,$ will simultaneously symmetrize the effective
conductivity too, (2) one parameter, say $c,$ is effectively
factorized and, since the conductivities "that differ only by a
proportionality constant are obviously similar", can be taken
equal to 1. These conjectures correspond, instead of (11), to the
following transformation of $\sigma_e$ (see also schematic
diagrams of different transformations in fig.1)
$$
T_m (\sigma_e(\{\sigma_i\},\{x_i\})) = \sigma_e( T_m
(\{\sigma_i\}),\{x_i\}), \quad T_m(z) = \fr{z-ib'}{-iz + a}. \en(11')
$$

\begin{figure}
\begin{picture}(250,120)
\put(65,33){\vector(1,0){50}}
\put(88,40){$T_m$}
\put(123,90){$\sigma'_e$} \put(63,90){\vector(1,0){50}}
\put(113,95){\vector(-1,0){50}}
\put(88,80){$T_m$}
\put(88,103){$T_m^{-1}$}
\put(120,30){$\{\sigma'_i\}$}
\put(40,30){$\{\sigma_i\}$} \put(45,90){$\sigma_e$}
\put(80,5){(a)} \put(130,0)
{\begin{picture}(80,50)%
\put(95,33){\vector(1,0){50}}
\put(118,40){$T_h$}
\put(118,83){$T_h$}
\put(157,45){\vector(0,1){40}}
\put(78,45){\vector(0,1){40}} \put(155,90){$\sigma'_e$}
\put(143,93){\vector(-1,0){50}} \put(150,30){$\{\sigma'_i\}$}
\put(70,30){$\{\sigma_i\}$} \put(75,90){$\sigma_e$}
\put(110,5){(b)}
\end{picture}}
\end{picture}

\vs{0.5cm}

{\small Fig.1. (a) A schematic diagram of transformations $T_m$ used in
\ci{16}, (b)  a schematic diagram of transformation $T_h$ used in this
paper.}
\end{figure}

Since it differs from (11), we will construct an analogous
transformation in an accordance with the transformation rules (11)
(fig.1b), taking into account all 3 parameters. The parameters of
such transformation (let us call it $T_h$) must satisfy two
following equations, denoting a cancellation of imaginary parts of
$\sigma'_i$ (i.e. $Im \; T(\sigma_i)=0$)
$$
\sigma_{it}' = \bar c \fr{ \sigma_{id}^2 + (\bar a + \sigma_{it})(\sigma_{it}
-b/c)}{(\sigma_{id})^2 + (\bar a + \sigma_{it})^2} =0, \; (i=1,2).
\en(36)
$$
Note, that this transformation is not equivalent to a usual
diagonalization of matrices $\hat \sigma_i,$ since they, due to
their antisymmetric parts, cannot have real eigenvalues. Two
equations (36) are not enough for a complete determination of
$T_h$ (i.e. the parameters $\bar a_h,\bar b_h,\bar c_h$, for
brevity, we will not write this additional index $h$ and bars in
this section) in terms of partial conductivities of the initial
system $\sigma_{id},\sigma_{it}.$ The equations (36) permit the
determination of the parameter $a$ and the ratio $b'=b/c$
$$
a_{\pm} = \fr{|\sigma_2|^2 - |\sigma_1|^2 \pm
\sqrt{B}}{2(\sigma_{1t} - \sigma_{2t})}, \quad b'_{\pm} =
\fr{|\sigma_1|^2 - |\sigma_2|^2 \pm \sqrt{B}}{2(\sigma_{1t} -
\sigma_{2t})},
$$
$$
B = [(\sigma_{1t} - \sigma_{2t})^2 + (\sigma_{1d} -
\sigma_{2d})^2] [(\sigma_{1t} - \sigma_{2t})^2 + (\sigma_{1d} +
\sigma_{2d})^2],
\en(37)
$$
where, evidently, $B\ge 0.$ It follows from (34) that their
denominators $D_i$ can be represented in the factorised form
$$
D_i = \sigma_{id}^2 + (a+ \sigma_{it})^2 = (a+\sigma_{it})(a+b'),
\quad i=1,2. \en(38)
$$
Then, the diagonal (or real) parts of $\sigma_i$  transform under
$T_h$ as
$$
\sigma_{id}' =  \sigma_{id} \fr{c(a + b')}{(\sigma_{id})^2 + (a+
\sigma_{it})^2} =   \fr{c \sigma_{id}}{\sigma_{ai}}, \quad \sigma_{ai} = a + \sigma_{it}. 
\en(39)
$$
The primed diagonal components are determined up to a common
factor $c$ (it was assumed to be 1 in \ci{16}.) In order to find
$c$ (or $b$) one needs one additional relation on parameters
$a,b,c.$ Since the transformation $T_h$ does not depend on phase
concentrations, it is convenient to choose an additional relation
by requiring that $T_h$ must reproduce one of the two exact
expressions for components of $\sigma_e$ at equal phase
concentrations $x_1=x_2=1/2$ from (21). For example, the
requirement of a reproduction of $\sigma_{ed}$ with a usage of the
diagonal components $\sigma_{id}'$ from (39) and the exact
expression for $\sigma_e'= \sqrt{\sigma_{1d}' \sigma_{2d}'}$ at
$x_1=x_2=1/2$ gives the desired additional relation
$$
A = \left[1+\left(\fr{\sigma_{1t} - \sigma_{2t}}{\sigma_{1d} +
\sigma_{2d}}\right)^2\right]^{1/2} = \fr{(a+b')(\sigma_{a1}
\sigma_{a2})^{1/2}}{\sigma_{1d} \sigma_{2d} + (a/c)^2 \sigma_{a1}
\sigma_{a2}}. 
\en(40)
$$
Using the relation (40) one can determine only  
$|c|$ through $A$ and $\sigma_{ai},$ which depend on partial
conductivities $\sigma_{id}, \sigma_{it},$
$$
c^2 =  \fr{a^2 A \sigma_{a1} \sigma_{a2}} {(a+b')(\sigma_{a1} \sigma_{a2})^{1/2} - \sigma_{1d} \sigma_{2d} A}.
\en(41)
$$
In order to find a sign of $c$ and to check the self-consistency of the transformation $T_h$ one must reproduce the exact expression for the transverse component $\sigma_{et}$ (and vice versa, if the requirement of a
reproduction of $\sigma_{et}$ is chosen as a first one). This means that the
following relation must hold
$$
\fr{\sigma_{1d} \sigma_{2t} + \sigma_{2d}
\sigma_{1t}}{\sigma_{1d} + \sigma_{2d}}= c\fr{{\sigma'_e}^2 -a
b'}{{\sigma'_e}^2 + a^2} =  c \fr{\sigma_{1d} \sigma_{2d} - (a
b'/c^2) \sigma_{a1} \sigma_{a2}}{\sigma_{1d} \sigma_{2d} + (a/c)^2
\sigma_{a1} \sigma_{a2}}. \en(42)
$$
The relations (40)-(42) give us a
highly nontrivial check of a self-consistency of the derived
transformation. The direct check of them is a rather complicated
task. Fortunately, there is more simple and elegant way to see
this, which can make the derived formulas more understandable. 

Let us consider the action of the transformation $T_h$ in the complex
plane $z$ in more details. By construction, it maps the given pair
of complex partial conductivities $\sigma_{1}, \sigma_{2},$ into the
real axis. According to general properties of the LFT, $T_h$ must map
into the real axis a whole circumference, containing partial
conductivities $\sigma_{1}, \sigma_{2}$ (see fig.2).

\begin{figure}[tb]
\cl{\input epsf \epsfxsize=8cm \epsfbox{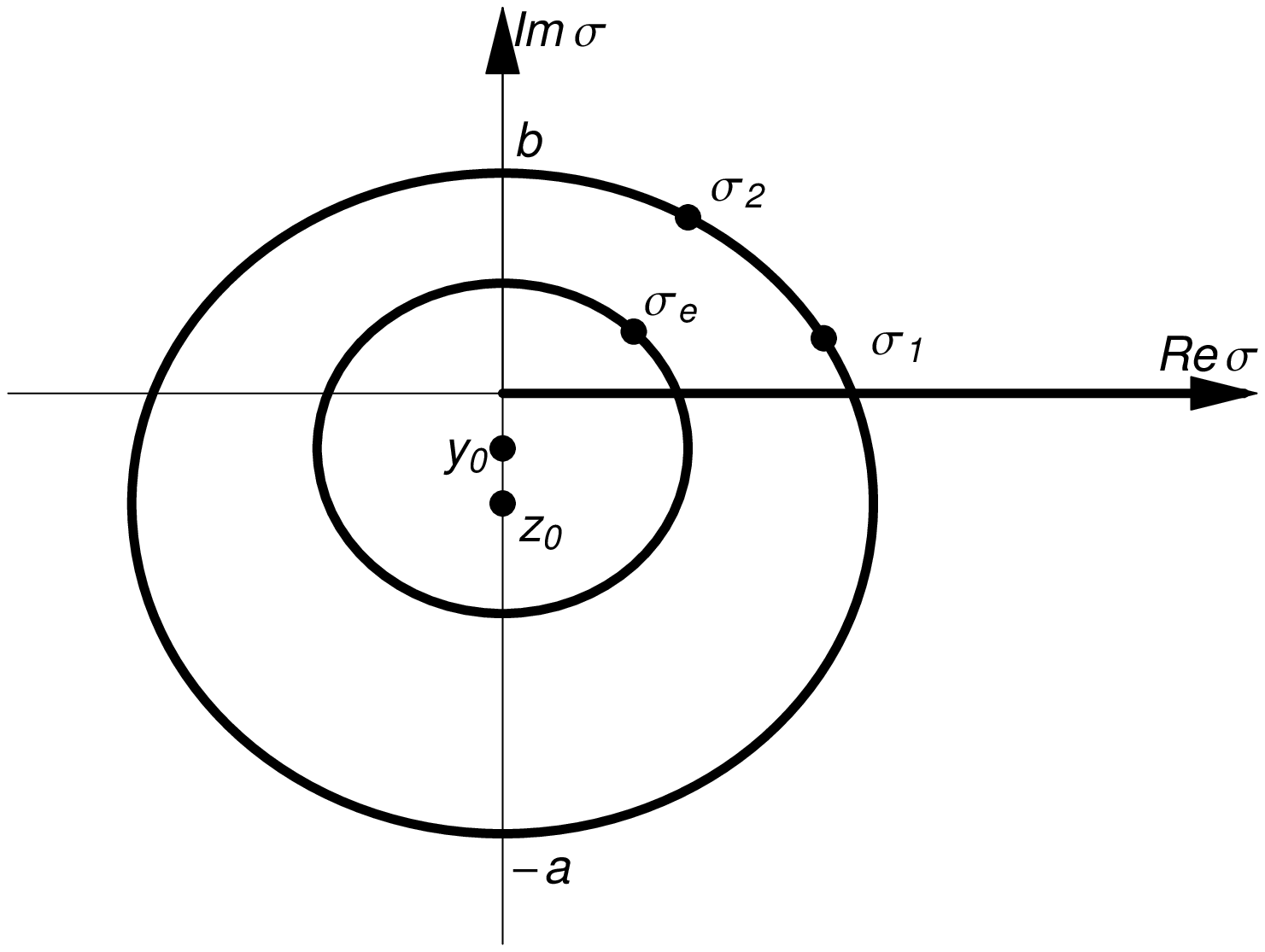}}

\vs{0.5cm}

{\small  Fig.2.  A schematic picture of two circles on the complex conductivity
plane, appearing under transformation $T_h$ used in this paper.}
\end{figure}
This fact was noted in \ci{16}. It allows to determine a position
of the center of this circumference, which is situated on the
imaginary axis in the point $z_0$, and its radius $R$
$$
z_0= (b'- a)/2 = \fr{|\sigma_1|^2 -
|\sigma_2|^2}{2(\sigma_{1t} - \sigma_{2t})}, \quad
R= |a+b'|/2 = \fr{\sqrt{B}}{2|\sigma_{1t} - \sigma_{2t}|}.
\en(43)
$$ 
At the same time $T_h$ maps the real axis into another circumference described by the equation
$$
x^2 + (y-y_0)^2 = \fr{c^2 R^2}{a^2}, \quad y_0 = \fr{c(a-b')}{2a} = -\fr{cz_0}{a}.
\en(44)
$$
Just into this circumference $T_h$ maps $\sigma'_e.$ Using the
explicit formulas for $\sigma_e$ from (21), one can check that
$\sigma_e$ {\it belongs to the first circumference.} In order for
$T_h(\sigma'_e) = \sigma_e,$ these two circumference must
coincide! This gives a more simple expressions for $c$ and $b$
$$
c = -a, \quad b = b'c = - a b'. \en(45)
$$
The equality $c = -a$ means also that effectively $T_h^2 \sim I,$
since it transforms the first circle into itself. 
Substituting this parameters in (39) one obtains the explicit expressions for
$T_h$ and $\sigma'_{id}.$ One can check also that the relations (40)-(42) are fulfilled (for both $a_{\pm}$). It is interesting to compare the inverse transformation $T_m^{-1},$ found in
\ci{16} under conjecture $c=1$,  with $T_h$ with
$c=-a,$ when they act on their primed conductivities, belonging to the real axes $y=0$ (it is obvious that they differ in general),
$$
T_m^{-1}(z) = \fr{az + ib'}{iz+1} \xrightarrow[y=0]{} \fr{(a+b')x
+i(-ax^2+b')}{1+x^2},
$$
$$
T_h (z) = \fr{az + ib'}{iz-a} \xrightarrow[y=0, x= -a \bar x]{}
\fr{(a+b')\bar x - i(a\bar x^2 -b')}{1+\bar x^2}, \en(46)
$$
where we have used the fact that the primed values for $T_h$ 
contain a common factor $c=-a,$ i.e. $x= -a \bar x.$ One can see that in this case they luckily coincide! 
Since $T_h (\sigma'_e)$ belongs at $c=-a$ to the first
circumference, this automatically ensures a fulfillment of the
second requirement. All equalities necessary for fulfillment of (39) correspond to the known equalities between lengths of various chords, their
projections on a diameter and the radius of the circles 
(see fig.3.). 
\begin{figure}[tbp]
\cl{\input epsf \epsfxsize=8cm \epsfbox{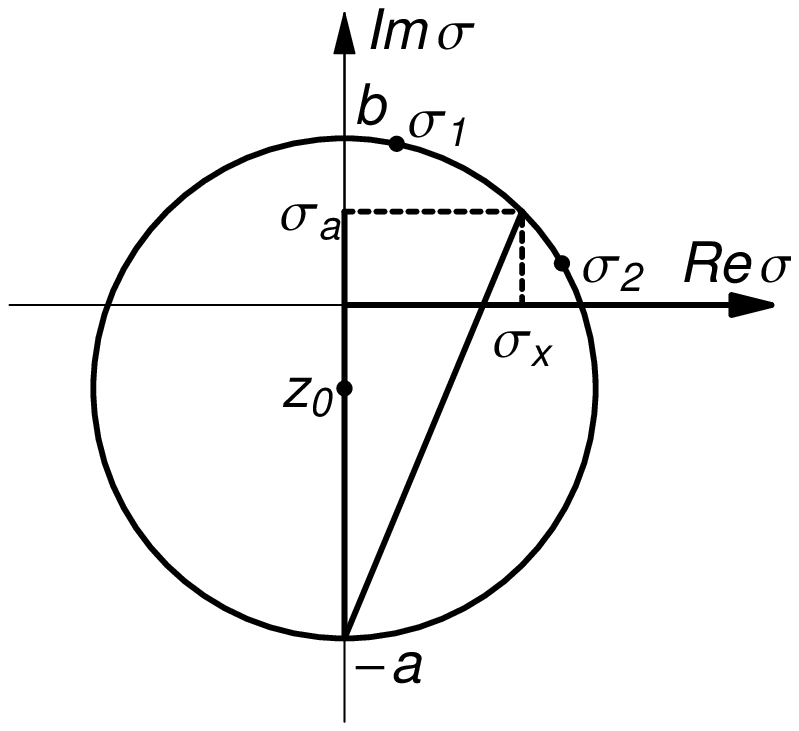}}

\vs{0.5cm}

{\small  Fig.3.  A schematic picture of the chord, defining a
geometrical sense  of the factor $A$ from the expression for the
real part of the exact $\sigma_e$ .}
\end{figure}
For
example, the equality (37) takes the form
$$
A= \fr{2R\sigma_a }{\sigma_a^2 + \sigma_{1d} \sigma_{2d}} = 1/(1-\delta/2R\sigma_a), \quad
\sigma_a^2 = \sigma_{a1} \sigma_{a2},
$$
where $\sigma_a^2$ is a squared geometrical average
of $a+\sigma_i, \; (i=1,2),$ and $\delta$ defines a  difference
between $\sigma_x^2 = 2R\sigma_a - \sigma^2_a,$ a squared real
projection of the chord, having an imaginary projection
$\sigma_a,$  and a squared geometric average of $\sigma_{id}$
$$
\delta = \sigma_x^2 - \sigma_{1d} \sigma_{2d}.
$$
It is worth to note that the transformation $T_h$ with $c=a$ transforms initial system into that with an opposite magnetic field $-{\bf H}.$

Thus, the obtained full (3-parametric) transformation $T_h$ allows
to find the effective conductivity of binary inhomogeneous system
in magnetic field at arbitrary phase concentrations in an explicit
form, if the partial conductivities of the initial system in
magnetic field $\sigma_{id}, \sigma_{it},$ and the effective
conductivity of artificial system at arbitrary phase
concentrations are known. This question together with other
methods of construction of explicit expressions for $\sigma_e$ in
magnetic field at arbitrary phase concentrations and their high
magnetic field behaviour will be considered in detail in the
forthcoming paper.

\bs
\und{5. Conclusion}
\bs

Using the exact duality transformations and symmetry properties of
2D isotropic heterophase systems we have found all possible physical
fixed points (except possible very special FPs with unusual properties)
and the corresponding exact values of their effective
conductivities in magnetic field. The obtained results take place
for various heterophase systems (regular and nonregular as well as
random), satisfying the symmetry and self-duality conditions, and
show very interesting properties. As in the case $N>2, H=0$, an
existence of hyperplanes in the concentration space with  constant
$\sigma_{ed,et}$, when partial conductivities belong to some
surfaces in their space, is unusual. These results admit a direct
experimental checking. The 3-parametric
transformation, connecting effective conductivities of two-phase
systems with magnetic field and without it, was also constructed.
It allows to find effective conductivity of two-phase systems  in
a perpendicular magnetic field at arbitrary phase concentrations,
if the effective conductivity of these systems at $H=0$ is known.

\bs
\und{ Acknowledgments}

\bs
The authors are thankful to Prof.A.P.Veselov for very useful
discussions of some mathematical questions. This work was
supported by the RFBR grants 00-15-96579, 02-02-16403, and by the
Royal Society grant 2004/R4-EF.

\end{document}